\documentstyle[preprint,aps]{revtex}

\begin{document}
\draft
\title{Effect of $\gamma$-irradiation \\
on superconductivity 
in polycrystalline YBa$_{2}$Cu$_{3}$O$_{7-\delta}$
}
\author{B. I. Belevtsev$^{a}$$^\star$, I. V. Volchok$^{b}$, 
N. V. Dalakova$^{a}$, V. I. Dotsenko$^{a}$, \\ 
L. G. Ivanchenko$^{a}$, A. V. Kuznichenko$^{c}$, I. I. Logvinov$^{a}$ 
}
\address{$^{a}$B. Verkin Institute for Low Temperature 
Physics \& Engineering, Kharkov, 61164, Ukraine\\
$^{b}$Kharkov State Technical University of Agriculture, 
Kharkov, 61002, Ukraine\\
$^{c}$National State University, Kharkov, 61077, Ukraine\\
}
\maketitle
\begin{abstract}
A bulk polycrystalline sample of YBa$_{2}$Cu$_{3}$O$_{7-\delta}$ 
($\delta \approx 0.1$) has been irradiated by $\gamma$-rays with $^{60}$Co source.
Non-monotonic behavior of $T_{c}$ (defined as the temperature at which normal
resistance is halved) with increasing irradiation dose $\Phi$ (up to about
220 MR) is observed: $T_{c}$ decreases at low doses ($\Phi \leq 50$~MR) from
initial value ($\approx 93$~K) by about 2 K and then rises, forming minimum.
At highest doses ($\Phi \geq 120$~MR) $T_{c}$ goes down again. The temperature
width, $\delta T_{c}$, of resistive transition increases rather sharp with
dose below 75 MR and somewhat drops at higher dose. We believe that this 
effect is revealed for the first time at $\gamma$-irradiation of high-$T_{c}$
superconductor. The cross sections for the displacement of lattice atoms in 
YBCO by $\gamma$-rays due to the Compton process were calculated, and possible
dpa values were estimated. The results obtained are discussed taking into
account that sample studied is granular superconductor and, hence, the 
observed variations of superconducting properties should be connected 
primarily with the influence of $\gamma$-rays on intergrain Josephson 
coupling.
\end{abstract}

\section{Introduction}
\par
The influence of crystal lattice disorder on superconducting properties is  
one of the important problems of superconductivity. Theoretical 
insights into this field have been initiated by seminal Anderson's 
work\cite{anders}. In low-$T_{c}$ superconductors (LTSCs) the 
disorder influence on critical temperature $T_{c}$ is believed to be 
mainly due to  dependences of density of states at the Fermi 
level $N(E_{F})$, constant of electron-phonon interaction  
$\lambda_{ep}$ or Coulomb interaction on disorder (see reviews in Refs.\ 
\onlinecite{bel1,sad,kim}). Many aspects of this problem are still not 
understood clearly. Things get much worse in the case of 
high-$T_{c}$ superconductors (HTSCs) which are layered cuprates 
with perovskite-related structure. Really, it is hardly 
possible to consider the crystal-disorder effects in HTSCs quite properly in 
circumstances where the nature of superconducting pairing in these 
compounds is still controversial \cite{ruv,ford}. 
Notwithstanding this handicap, the crystal-lattice disorder effects in 
HTSCs have become topic of a large body of experimental and theoretical 
research\cite{sad,kim}. The necessity of such type of studies is quite 
obvious from technological and fundamental points of view.
\par
In HTSCs as well as in LTSCs the superconductivity involves pairs of 
electrons. Therefore, although the nature of superconducting paired state 
in the HTSCs is still obscure, it can be assumed that some features of 
the disorder influence should be common for both types of superconductors. 
This can be seen from the following consideration. The order 
parameter in the Ginzburg-Landau theory is written as 
$\Psi = \Delta \exp(i\varphi)$ where $\Delta$ is the amplitude and $\varphi$ 
the phase of this parameter. 
The disorder can destroy the superconductivity 
either by reducing the amplitude of the order parameter or by destroying
the phase coherence of superconducting electrons. The relative contribution 
of each of these mechanisms depends on the degree of inhomogeneity of the 
superconducting system\cite{bel1}. The role of inhomogeneity can be judged by
comparing the two characteristic lengths $\xi_{d}$ and $\xi_{GL}(T)$, 
the first of which is characteristic scale of inhomogeneity associated with 
disorder and the second is the Ginzburg-Landau coherence length for dirty
superconductors
$$
\xi_{GL}(T) = \xi_{GL}(0)\epsilon^{-1/2}, \eqno(1)
$$
\noindent where $\xi_{GL}(0) = \pi D\hbar/8kT_{c}$ is the coherence length 
at $T=0$, $D = v_{F}l/3$ is the electron diffusion coefficient 
($v_{F}$ is Fermi velocity, $l$ is electron elastic scattering length), 
$\epsilon = \ln(T_{c}/T$).  When 
$\xi_{d} \ll \xi_{GL}(T)$ (as for instance, in the case of the impurity or 
point defects of crystal lattice), the system behaves as homogeneous one, 
and disorder affects mainly the order parameter.  In the opposite case, 
when $\xi_{d} \gtrsim \xi_{GL}(T)$, the system is inhomogeneous. An example 
of such system is granular metal consisting of metal grains separated by 
dielectric interlayer ( $\xi_{d}$ is grain size in this case). 
The reduction in the critical temperature $T_{c}$ in inhomogeneous systems
is largely due to the suppression of phase coherence between weakly coupled 
superconducting regions. Both mechanisms of the $T_{c}$ reduction under 
disorder (reducing 
the amplitude of the order parameter and  destroying the phase coherence 
of superconducting electrons) can operate simultaneously in real systems. 
\par
The HTSCs are usually characterized by low values of $v_{F}$ and $D$ as 
compared with that of LTSCs. Together with high $T_{c}$ values, this should 
lead, according Eq.\ (1),  to exceptionally short superconducting coherence 
length. Indeed, it was found, for example, in YBa$_2$Cu$_3$O$_{7-\delta}$
that $\xi_{GL}(0)$ is 
about $\xi_{ab} \approx 1.4$~nm and $\xi_{c} \approx 0.2$~nm, parallel and perpendicular to the 
CuO$_{2}$--layer\cite{ford}. From this it follows that the influence of 
structure inhomogeneity and thus the suppression of phase coherence should 
be in general far more important in HTSCs than in LTSCs. 
\par
Actually the HTSC films or bulk samples are always disordered and 
inhomogeneous to some extent.  For example, in polycrystalline samples the 
regions of grain boundaries  are highly disordered. These regions can be not 
only non-superconductive, but dielectric as well\cite{boyko,stolb}. In fact, 
polycrystalline samples are granular metals in which superconducting 
coherence between the grains can be established by Josephson coupling.
Beside the granularity, chemical-composition inhomogeneity 
is quite common in HTSCs. The influence of this type of inhomogeneity 
on superconductivity can be rather strong since charge carrier density 
of HTSCs depends strongly on chemical composition\cite{ruv,ford}. 
The composition disorder can contribute to destroying of superconducting 
phase coherence in the same fashion as the granularity. In most cases the 
inhomogeneous distribution of oxygen leads to this type of disorder.
Two main sources for oxygen inhomogeneity can be distinguished: extrinsic 
and intrinsic. Extrinsic source is due to various technological factors
of sample preparation. The intrinsic one is connected with phase separation 
of HTSCs on two phases with different oxygen and hence charge carrier 
concentration\cite{ps}. 
\par
It is clear from the aforesaid that to a good approximation the two main 
types of disorder, which are essential for superconductivity, can be
distinguished. The first of them is disorder associated with perturbations 
of crystal lattice on the atomic scale (impurities, vacancies, and the like). 
This type of disorder is often called microscopic. It can be
responsible for electron localization and other phenomena\cite{bel1,sad,kim} 
which affect the superconducting order parameter. The second type of 
disorder is
associated with structural inhomogeneity of superconductor (granular 
structure, phase separation, inhomogeneity due to technological reasons).
The disorder scale in this case is much more than interatomic 
distances (at least, more than, say, 10-100 nm) and hence, this disorder 
can be called macroscopic. The macroscopic disorder affects mainly the 
superconducting phase coherence. In experimental studies it is desirable 
to separate the effects of microscopic and macroscopic disorder or,
at least, to be aware of the possible joint action of them in the case
where such separation is difficult. Ignoring this point could lead 
to serious errors in interpretation of experimental results or to 
total their misunderstanding. 
\par
In study of disorder effects in HTSCs it is important to use reliable 
methods of controllable disordering.  Ideally, it is desirable to "tune" 
disorder while keeping other parameters fixed.  
At present, the main method which is expected to be somewhat close to such 
ideal is irradiation of superconductor with fast (charged or uncharged) 
particles or with high-energy photons ($\gamma$-rays). 
By appropriate  choice of particles and their energy  
the irradiation would be expected to produce only the microscopic
disorder (vacancies and interstitials), although in the HTSCs a 
possibility of inducing of compositional disorder must be always taken 
into account. 
\par
In this communication we present some new results of investigation 
of influence of $\gamma$-irradiation on the critical temperature of bulk 
polycrystalline YBa$_2$Cu$_3$O$_{7-\delta}$ (YBCO). The disordering of HTSCs
with $\gamma$-rays was used rather often in previous experiments, especially 
in the first years after discovery of 
HTSCs\cite{bohandy,kato,boi,kutsu,vasek,polyak,ozkan,rangel}. 
Such type of investigations could have an applied importance in the case of 
possible use of HTSCs in environments of nuclear reactors or space stations. 
The fundamental importance of experiments of this sort is also beyond 
question. 
\par
It is known that $\gamma$-rays produce ionization in 
solids\cite{bethe,dienes}. In metals, ionization produced by radiation is 
very rapidly neutralized by the conduction electrons\cite{dienes}. Beside 
this, $\gamma$-rays produce displaced atoms in solid, namely, 
vacancy-interstitial pairs (Frenkel defects) of small separation, randomly 
distributed through the lattice. Interaction of $\gamma$-rays with matter, 
which leads to atomic displacements, occurs principally by means of three 
mechanisms: the photoelectric effect, the Compton effect, and pair 
production\cite{bethe,dienes}. In all three processes electrons are 
ejected with energy comparable with the original $\gamma$-ray energy, 
and thus $\gamma$-ray irradiation inevitably causes a substance to be 
internally bombarded by fairly energetic electrons\cite{dienes}. 
At $\gamma$-ray energy of a few MeV, the dominant contribution to 
displacement production comes from Compton effect. 
\par
When compared with other irradiation sources, $\gamma$-rays have one 
unquestionable advantage. Attenuation distances of $\gamma$-rays with energies 
of few MeV are order of a few centimeters. This enables one to investigate 
bulk samples of HTSCs.  By contrast, penetration distances for irradiation 
with particles or ions are small. In this case it is possible to 
study the disorder effects quite properly only in rather thin films of HTSCs. 
Otherwise, it can not be excluded the implantation and doping effects as 
well as inhomogeneity of damage that plagues the interpretation of the 
experimental results.
\par
A disadvantage of $\gamma$-rays is that the available monoenergetic 
radiactive sources give low $\gamma$-ray fluxes. Taking into account 
the effective cross section for atomic displacement through the Compton 
mechanism\cite{dienes} one would not expect any significant changes in the 
conductivity of typical metals, like copper. Indeed, a typical dose of 
$\gamma$-ray irradiation is about $1\times 10^{17}$~photons/cm$^{2}$. 
For this dose, the number of displacements per atom (dpa) for copper is about 
$10^{-8}$\cite{dienes,kinchin}. The calculated increase in resistivity of
copper due to 1\% of point defects (that is for 0.01 dpa) is about 
1~$\mu\Omega$~cm\cite{dienes,kinchin,seitz}. From this it is apparent that 
the fraction of displaced atoms, which can be produced through the Compton 
mechanism in a convenient irradiation time, should not induce any noticeable 
resistance variations in good metals.  However, in poor conductors with  
low charge density, such as semiconductors, $\gamma$-rays can strongly 
influence the conductivity through the carrier removal\cite{oen,cahn}.       
\par
Charge-carrier density in optimally doped HTSCs (such as YBCO) is usually 
about $5\times10^{21}$~cm$^{-3}$\cite{sad}, that is well below that of 
good metals. This density is not however low enough to expect some 
significant influence of $\gamma$-ray irradiation on conductivity and 
order parameter of HTSCs, and hence, on their $T_{c}$. In spite of this,
in some known experiments a quite appreciable  effect of $\gamma$-irradiation 
on $T_{c}$ and resitivity of YBCO was revealed. The maximal decrease in 
$T_{c}$ was found to be about 2K by high doses $\Phi \simeq 
1000$~MR\cite{kato,vasek}. In general, the published 
experimental results\cite{bohandy,kato,boi,kutsu,vasek,polyak,ozkan,rangel} 
about the effect of $\gamma$-irradiation on YBCO are quite contradictory 
(all of them concern the optimally doped YBCO with $T_{c}$ above 90 K). 
For example, in contrast to Ref.\ \onlinecite{kato,vasek},  in some other 
investigations no change in $T_{c}$ was found at all up to doses about 
1000 MR\cite{kutsu}, or found to be much less than 1 K\cite{ozkan,rangel}.
The results about $\gamma$-ray effect in resistivity are even more 
conflicting. It is possible to find among them such extremal cases:
a) No influence at all up to dose $\Phi \simeq 1000$~MR\cite{kutsu};
b) A nearly tenfold increase in resistivity at 
$\Phi \simeq 150$~MR\cite{rangel}. 
\par
The above-mentioned relevant $\gamma$-ray 
papers\cite{bohandy,kato,boi,kutsu,vasek,polyak,ozkan,rangel} just present
experimental results without much (or any) consideration of possible 
mechanisms of $\gamma$-ray influence on conductivity and $T_{c}$ of HTSCs, 
or presentation of general view on such influence. In none of them the 
degree of radiation damage (the value of dpa) has been estimated that
severely hinders the understanding and interpretation of the results. 
It may be inferred, therefore, that the undestanding of $\gamma$-ray 
influence on HTSCs is far from completion and that further experimental and 
theoretical investigations of this matter are necessary.
\par
In this paper, we report investigation of the effect of $\gamma$-rays on
$T_{c}$  and resistive transition to superconducting state in bulk 
polycrystalline sample of optimally doped YBCO. We have observed  
non-monotonic behavior of $T_{c}$ with increasing irradiation dose $\Phi$ 
(up to about 220 MR): $T_{c}$ decreases at low doses ($\Phi \leq 50$~MR) 
approximately by 2 K and then rises, forming a minimum. Quite unexpected, 
at highest doses ($\Phi \geq 120$~MR) $T_{c}$ value goes clearly down again. 
The temperature width, $\delta T_{c}$, of resistive 
superconducting transition increases rather sharp with dose in the range
$\Phi \leq 75$~MR and somewhat drops at higher dose. We believe that this interesting 
effect is revealed for the first time. Unlike previous 
works\cite{bohandy,kato,boi,kutsu,vasek,polyak,ozkan,rangel}, we have 
calculated cross sections for the displacement of different kinds of lattice 
atoms in YBCO by $\gamma$-rays due to the Compton process. To our knowledge, 
nobody has done it before for HTSCs. This  enables us to estimate the possible 
dpa value and has facilitated the interpretation of results. 
\par
On the strength of the obtained results and those of previous investigations, together with the results of radiation damage calculation, 
we come to conclusion that for commonly used doses (up to 
$\Phi \simeq$~1000 MR) $\gamma$-rays should not produce any substantial 
influence on superconducting properties and conductivity of single-crystal 
HTSCs.  In polycrystalline samples however this influence can be fairly 
strong. These samples, as was mentioned above, are in fact granular metals 
which consist of metallic grains separated by dielectric interlayers. 
In the HTSCs the regions of grain boundaries, and near environtments of 
them, are strongly depleted with charge carriers\cite{boyko,stolb} and thus 
should be very sensitive to $\gamma$-rays and radiation of any other kind.  
For this reason the conductivity and $T_{c}$ in polycrystalline HTSCs can be 
markedly affected by $\gamma$-rays even at commonly used not very high doses. This is the main 
conclusion of our paper. Based on this it is possible to 
understand, at least qualitatively, the sharp dictinction between the known 
published results about $\gamma$-ray influence on YBCO. In this paper, the 
possible mechanisms of impact of radiation-damage in grain boundary regions 
upon the conductivity  and $T_{c}$ in YBCO are considered.  Beside the atomic 
displacements, the probable effects of atom ionization in grain boundary
regions by $\gamma$-rays are speculatively discussed.

\section {Sample and experiment} 

For sample preparing the conventional solid-state reaction method
was used. Fine powders of Y$_{2}$O$_{3}$, BaO and CuO were mixed in alumina 
mortar in the composition of YBa$_2$Cu$_3$O$_{7-\delta}$. Then the mixture
was pressed to pellet at pressure about 100 MPa. This pellet was heated at
$T\approx 930$$^{\circ}$C in air and then slowly cooled down. From the 
pellet a 3.1$\times$6.3$\times$22~mm$^{3}$ bar was cut. The oxygen content in 
sample was estimated to be about 6.9 ($\delta \approx 0.1$). The sample 
was polycrystalline with rather large grain size (about 12 $\mu$m). 
Density of sample was 5.2 g/cm$^{3}$, that comprises about 80\% of 
the expected density for this compound. The 20\% difference can
be attributed to an occurence of small voids and pores that is quite usual 
for polycrystalline samples.
\par
Irradiation was carried out using $^{60}$Co $\gamma$-ray source at room 
temperature in air at different doses up to $\Phi \approx 220$~MR. This
isotope emitts the equal amounts of photons with energies 1.17 MeV and 
1.33 MeV. The dose rate of used source was $6.34\times10^{4}$~R/h. 
The temperature dependences of resistance $R(T)$ were measured by standard
four-probe method with a direct measuring current 10 mA. The accuracy of 
temperature measurements was estimated to be about 0.1 K. 

\section{Radiation-damage calculation}
\label{sect1}
In experimental investigations of radiation effects in solids it is 
imperative  to realize the degree of disorder produced by irradiation in 
objects studied. Without such knowledge, adequate understanding and 
interpretation of results obtained is impossible. In many cases such 
ignorance can lead to wrong conclusions. For this reason, calculations of 
possible radiation damage, primarilly the dpa values, become nowadays an 
integral part of the most of radiation-effect investigations. For example, 
at ion irradiation studies the well-known TRIM simulation 
program\cite{ziegler} is widely used. The surprising thing is that one 
cannot find even rough estimates of dpa in the known studies of 
$\gamma$-ray effect in 
YBCO\cite{bohandy,kato,boi,kutsu,vasek,polyak,ozkan,rangel}. 
This type of calculations is rather cumbersome and may be for this reason
they had not been done in the above-mentioned studies. The principal concepts of 
such calculations are however quite established\cite{bethe,dienes,oen,cahn} 
and therefore the estimation of dpa or concentration of point defects 
induced by $\gamma$-rays can be done without principal difficulties. 
\par 
In this work we have calculated the cross sections for atomic displacements 
in YBCO by $\gamma$-rays due to the Compton process which is a main producer 
of energetic electrons at photon energy about 1 MeV\cite{bethe,dienes}.   
Somewhat different versions of this type of calculation are presented in
Ref.\ \onlinecite{dienes,oen,cahn}. In this study  the version of Ref.\ 
\cite{dienes} was used. The number of atoms displaced in unit volume 
(cm$^{3}$) per second is expressed in Ref.\ \onlinecite{dienes} by
$$
R_{d}^{\beta} =  
\int_{0}^{E_{max}}n_{0}\: \sigma_{d}^{\beta}(E)\: \Phi_{\beta}(E)\: dE, 
\eqno(2)
$$
\noindent where  $n_{0}$ is number of atoms per unit volume; 
$\sigma_{d}^{\beta}(E)$ is cross section for an atom to be displaced by an
electron of energy $E$; $\Phi_{\beta}(E)$ is flux density of ejected Compton
electrons per cm$^{2}$/sec, at energy $E$, per unit range of energy;  
$E_{max}$ is maximal energy of Compton electron, which is given by 
$E_{max} = 2E_{\gamma}/(1+2E_{\gamma})$ with $E_{\gamma}$ being photon 
energy in monoenergetic $\gamma$-ray flux.  The Eq.\ (2) can be written in
more detail:
$$
R_{d}^{\beta} =  \Phi_{\gamma}\, n_{0}^{2} 
\int_{0}^{E_{max}}dE \: \sigma_{d}^{\beta}(E)
\: (-dE/dx)^{-1}\: \int_{E}^{E_{max}}\sigma_{c}(E^{'})\: dE^{'}, 
\eqno(3)
$$ 
\noindent where $\Phi_{\gamma}$ is $\gamma$-ray flux per cm$^{2}$/sec,
$\sigma_{c}(E^{'})$ is cross section per atom for the production of a
Compton electron at energy $E$. This is given by the formula\cite{bethe}
$$
\sigma_{c}(\epsilon) =  
\sigma_{0}\, \left\{ 
\frac{1}{1-\epsilon} + 1 - \epsilon + 
\frac{\epsilon}{\gamma^{2}(1-\epsilon)} \left [ \frac{\epsilon}{1-\epsilon} 
- 2\gamma \right ]
\right\}, 
\eqno(4)
$$
\noindent where $\epsilon = E/E_{\gamma}$, $\gamma = E_{\gamma}/mc^{2}$, and
$\sigma_{0} = \pi r_{0}^{2}\: Z_{2}\: mc^{2}/E_{\gamma}^{2}$. Here $e$ and
$m$ are electronic charge and mass, respectively, $c$ is velocity of light,
$r_{0} = e^2/mc^2$, and $Z_{2}$ is the atomic number. The energy loss per cm 
of electron path, $-dE/dx$, is given by\cite{dienes}
$$
-dE/dx =  
a\left [ \left ( 1 + E/mc^2\right )/E\right ] 
\eqno(5)
$$
\noindent where $a = 2\pi e^{4}n_{0}Z_{2}L$, $L \approx 10$. 
\par
The cross section $\sigma_{d}^{\beta}(E)$ in Eqs.\ (2) and (3) for an 
atom to be displaced by an electron of energy $E$ depends essentially 
on the specific value of the threshold energy $E_{d}$. It is 
assumed\cite{dienes,seitz} that an atom is always displaced from its 
lattice site when it receives energy greater then $E_{d}$ and is never 
displaced at lower energy. Therefore, it should be put down 
$\sigma_{d}^{\beta}(E) = 0$ in Eq.\ (3) if $E < E_{d}$. For the greater
values of $E$ we have used the known McKinley-Feshbach 
formula\cite{dienes,mckin} to calculate $\sigma_{d}^{\beta}(E)$:
$$
\sigma_{d}^{\beta}(E) =  
\frac{\pi}{4}b^2\left [ \left ( \frac{T_{m}}{E_{d}} -1 \right ) -
\beta^{2}\ln\frac{T_m}{E_d} +
\pi\alpha\beta\left \{ 2\left [ \left ( \frac{T_m}{E_d} \right )^{1/2} -1 
\right ] - \ln\frac{T_m}{E_d} \right \} 
\right ] 
\eqno(6)
$$
\noindent where $(\pi/4)b^2 = \pi Z_2 r_{0}^{2}(1-\beta^{2})/\beta^{4}$;
$\beta = v/c$ that is the ratio of electron and light velocities for which 
the relation $\beta^{2} = E(E+2)/(E + 1)^{2}$ is true if $E$ is taken in 
the units of $mc^2$; $\alpha \approx Z_{2}/137$, and $T_{m}$ is the maximum 
energy which can be transferred in a collision by an electron of kinetic 
energy $E$:
$$
T_{m} = \frac{2(E + 2mc^2)}{M_{2}c^{2}}E,
\eqno(7)
$$  
\noindent where $M_{2}$ is the target atomic mass. 
\par
Some additional comments about the McKinley-Feshbach formula are necessary.
It is assumed that it is accurate to one percent for $Z_{2}$ 
up to 40\cite{cahn}. Atomic numbers of elements in YBCO are 8(O), 29 (Cu),
39 (Y), and 56 (Ba). Therefore, an error more than one percent should be 
expected only for Ba atoms. But displacement cross section for heavy atoms is 
usually much less  than for light ones, and thus, contribution from heavy 
atoms to total dpa should be rather small.  For this reason we did not expect 
a significant error with the McKinley-Feshbach formula. This was justified 
by our calculations which are presented below.
\par
If the struck atom has large enough energy it can cause secondary 
displacements\cite{kinchin} which can also be taken into account at 
calculation of numbers of the displaced atoms due to 
$\gamma$-rays\cite{cahn}. It is easy to see however that at electron energy 
about 1 MeV the values of $T_{m}$ given by Eq.\ (7) are quite small and 
therefore this effect can be neglected. Nontheless, we have done cascade 
calculations following the recommendations of Refs.\ \onlinecite{kinchin,cahn} 
and have found that primary displacement cross 
sections differ from these of total displacement cross sections only by 
few percents. Although we will present below the total displacement cross 
sections, it should be kept in mind that they differ very slightly from 
those of obtained using Eqs.\ (4)--(7).
\par
The calculated values of the effective cross section for atomic 
displacement  by $\gamma$-rays from $^{60}$Co source through the Compton 
effect, $\sigma^{\gamma}_{c} = R_{d}^{\beta}/(\Phi_{\gamma}n_{0}$), at 
different values of threshold energy $E_{d}$ for ions in YBCO are given in
Table\ \ref{table1}. The values of $\sigma^{\gamma}_{c}$ presented in this
Table are the weighted averages for photons with energies 1.17 MeV and 
1.33 MeV which are emitted by $^{60}$Co source. The values of $n_{0}$  
for different ions at calculation of $\sigma^{\gamma}_{c}$ were just 
partial ion densities for compound YBa$_{2}$Cu$_{3}$O$_{7-\delta}$ 
deduced taking into account the compact YBCO mass density which is about 
6.4~g/cm$^{3}$. It can be seen from Table\ \ref{table1} that 
the $\sigma^{\gamma}_{c}$ values depend crucially on threshold energy $E_{d}$.  
It is worth noting also that  the $\sigma^{\gamma}_{c}$ values for 
heavy Ba ions are the least as compared with other ions, as expected. 
The data of Table\ \ref{table1} will be used below for estimation of 
possible dpa values in studied YBCO after $\gamma$-irradiation.     
 
\section{Results and discussion}

The temperature dependence of sample resistivity $\rho (T)$  before 
$\gamma$-irradiation is presented in Fig.~1. The dependence is quite 
common for optimally doped YBCO: it is linear above the 
superconducting resistive transition up to room temperature, 
$T_{c}$ value is about 93~K. We have defined experimental $T_{c}$ to be the
temperature at which normal resistance $R_{n}$ is halved. The dependences 
$R_{n}(T)$ in the range of resistive transition have been obtained by 
extrapolation of linear $R(T)$ dependence to this region. We have used the 
temperature $T_{cz}$ at which resistance goes to zero as a second 
characteristic of resistive transition. In fact, $T_{cz}$ is the end point 
of resistive transition to superconducting state, and is an essential 
characteristic which should be taken into account for granular or 
inhomogeneous superconductors. The difference in $T_{c}$ and $T_{cz}$,
$\delta T_{c} = T_{c} - T_{cz}$, is some quite definite measure of the 
width of resistive transition. 
\par
In early days of HTSC investigations (and quite often up to date) the 
experimental $T_{c}$ was defined as the temperature $T_{cb}$ at the onset 
of the superconducting transition.  Although this is also quite essential 
characteristic of resistive transition, it can be evaluated with much 
less precision than $T_{c}$ or $T_{cz}$. For this reason only 
the observed changes in $T_{c}$ and $T_{cz}$ with $\gamma$-ray dose will be 
considered below. 
\par
The $\rho(T)$ curves  for different $\gamma$-ray doses $\Phi$ in the 
temperature range of superconducting transition are presented in Fig.~2. It 
can be seen that $T_{c}$ depends on $\Phi$ in a non-monotonic way. More 
clearly this is seen in Fig.~3 where the changes in $T_{c}$ and $T_{cz}$ with  
$\gamma$-ray dose are shown.  It follows from the Figure that $T_{c}$ 
decreases at low dose ($\Phi \leq 50$~MR) by $\approx 2$~K and then rises, 
forming a minimum. At higher dose ($\Phi \geq 120$~MR) 
$T_{c}$ value goes clearly down again. The zero-resistance temperature 
$T_{cz}$ has been changing with $\gamma$-ray dose in the nearly same way as 
$T_{c}$, but with greater amplitude: the initial decrease in $T_{cz}$ is 
about 4 K. The magnitude of $\delta T_{c}$ (which characterizes 
the width of resistive transition and, hence, the sample inhomogeneity) 
increases with dose in the range $\Phi \leq 75$~MR and somewhat drops at 
higher doses (Fig.~4). 
\par
The observed initial $T_{c}$ decrease with $\gamma$-ray dose (Fig.~3) 
corresponds to some of previous studies\cite{kato,vasek}, but the general 
picture of non-monotonic dependence of $T_{c}$ on radiation dose looks like 
a surprising thing. To our knowledge such behavior of $T_{c}$ in HTSC 
with $\gamma$-ray dose is found for the first time. A somewhat (or partially)
similar  non-monotonic behavior has been seen previously \cite{boi} in 
$\gamma$-irradiated YBCO-related compound 
Y$_{0.9}$Sm$_{0.1}$Ba$_{2}$Cu$_{3}$O$_{7-\delta}$, but for the onset 
temperature $T_{cb}$ only. As this took place, the zero-resistance 
temperature $T_{cz}$ had not manifested any marked influence of 
$\gamma$-irradiation\cite{boi}. The resistive-transition curves shown in 
Ref.\ \onlinecite{boi} are rather steep (no large difference between the 
$T_{c}$ and $T_{cz}$ values). At the same time the reported values of the
onset temperature $T_{cb}$ (about 140~K) appear to be too high for 
YBCO-related compounds (compare, for example, with the data of Ref.\ 
\onlinecite{kutsu}). In our opinion in evaluating the onset temperature 
$T_{cb}$ from $R(T)$ curves the significant error is possible. Since an 
employed procedure for evaluating of the $T_{cb}$ values has not been 
outlined in Ref.\ \onlinecite{boi} one should take the $T_{cb}$ values 
in it and the described non-monotonic behavior $T_{cb}$ with 
$\gamma$-ray dose with some precaution. 
\par
The effect of $\gamma$-rays in the resistivity of sample studied was found
to be appreciable only in the temperature range of rather broad resistive 
superconducting transition (Fig.~2). With increasing temperature away from 
$T_{c}$ the $\gamma$-ray effect in resistivity falls off quickly. In 
particular, at $T \simeq 105$~K a mere 4\% increase in resistivity by 
$\gamma$-rays was found, whereas above 200 K hardly any radiation
effect in resistivity can be detected. This suggests that $\gamma$-rays 
affect primarily the superconducting properties of the sample studied
while  electron transport in normal state remains actually not affected.
\par
In the course of explaining of the results obtained, it should be taken 
into account, first of all, that the sample studied is polycrystalline 
or, better to say, granular. Indeed, its resistivity in normal state 
just above the superconducting transition (about 2.5 m$\Omega$~cm, as can 
be seen in  Figs. 1 and 2) is by a factor of 50 larger than that of the 
best quality optimally doped YBCO single-crystals   
(about 50 $\mu \Omega$~cm\cite{giap,tolp}). The increased resistivity comes 
from grain boundaries, since regions of grain boundaries in HTSCs can be poor
conductive and even dielectric\cite{boyko,stolb}.
\par
The conductivity of granular metals in normal state is determined by 
tunneling of single-particle excitations (unpaired electrons) through 
the boundaries. In superconducting state the superconducting coherence 
between grains can be established by Josephson coupling. Although the sample
studied is inhomogeneous (granular) its critical temperature $T_{c} = 93$~K 
corresponds to the highest $T_{c}$ values in the good-quality YBCO single 
crystals\cite{giap,tolp}. Such situation is quite possible and understandable 
for inhomogeneous systems\cite{bel1}. Granular metals usually have a spatial 
distribution of thicknesses of the poor-conductive or dielectric grain 
boundaries. This spread in the boundary thickness may be very important 
for transpor properties, especially for dielectric grain boundaries because 
in this case the probability of tunneling is an exponential function of 
dielectric separation distance. From this follows that granular metals 
are actually percolating systems\cite{deut}. In such systems 
conductivity can be determined by the presence of optimal ``chains'' of 
grains with maximal probability of tunneling for adjacent pairs of grains
forming the chain. The same type of percolation picture is true for a system 
of metallic grains with Josephson coupling between them  (when grains become 
superconducting at low enough temperature)\cite{deut}. The sample studied 
consists of rather large (12 $\mu$m) grains with critical temperature 
$T_{c}$ as high as that of in high-quality crystals.  
The global critical temperature of whole sample can be very close to this
maximal value in the presence of good conducting optimal chains of grains 
with strong Josephson coupling (in utmost case, theoretically, even one such 
path will be enough to arrive to zero resistance). 
\par
Since conductivity of the granular metal is determined by both intragrain 
and intergrain transport properties, the influence of radiation damage in 
it should be considered separately for materials of grains and regions of 
grain boundaries. Let us consider at first a question: is the radiation 
damage induced by $\gamma$-ray in this study high enough to cause quite 
significant variations in resistivity and superconducting properties of 
material of grains. For this purpose, we have estimated the 
possible values of dpa in sample studied using the calculated in 
Sec.\ \ref{sect1}  effective cross sections for atomic displacement by 
$\gamma$-rays from $^{60}$Co source through the Compton effect 
(Table\ \ref{table1}). It is easy to see that for the stardard value of
$E_{d} = 20$~eV  the total dpa comprises very small value about $10^{-7}$. 
There is experimental evidence that the displacement energy, $E_{d}$,  
for oxygen in CuO$_{2}$ planes in YBCO is close to 10 eV\cite{tolp}. 
In this case the dpa will be about $6\times 10^{-7}$. From the general 
point of view (taking into account the rather high charge-carrier density 
in optimally doped YBCO) it should not be expected any significant 
variations in normal state resistivity and $T_{c}$ in grains at such small 
radiation damage. Indeed, the known experimental studies of single-crystal 
HTSCs irradiated with electrons or ions show that an appreciable influence 
of disorder on resistivity and $T_{c}$ can be detected only at dpa greater 
than $\simeq 10^{-3}$\cite{giap,rull,vobornik}. 
\par
It follows from the above consideration that with commonly used  $\gamma$-ray 
doses ($\simeq 1000$~MR) one should not expect any detectable variations in 
resistivity and $T_{c}$ in fairly homogeneous single-crystal HTSC as well 
as in metallic grains of polycrystalline HTSC. For this reason we will not
consider here the numerous theoretical models of disorder influence on 
superconducting properties of HTSCs (see Refs.\ \onlinecite{sad,kim} and 
references therein) which were developed for homogeneous superconductors.
In this study we have not seen any marked changes even in the normal state 
resistivity of the polycrystalline sample. But we observed the pronounced 
effect of $\gamma$-rays on $T_{c}$ and the width of resistive transition 
(Figs. 2-4). This is undoubtedly connected with influence of $\gamma$-rays 
on poor conductive or dielectic regions of grain boundaries and, therefore, 
on the superconducting phase coherence between the grains.  
\par
One of the first impressive demonstration of radiation-induced destruction 
of phase coherence of the superconducting wave functions between grains in 
polycrystalline HTSC  was demonstrated in Ref.\ \onlinecite{clark} for 
optimally doped YBCO. They found that zero-resistance temperature $T_{cz}$ 
was very sensitive to even low doses of irradiation with 500 keV oxygen. 
At the same time, the onset temperature $T_{cb}$ declined much slowly. For 
doses, where $T_{cz}$ was already close to zero, and furthermore, even, 
when insulating behavior of 
$R(T)$ was evident below 40 K, $T_{cb}$ was close to 80 K (before 
irradiation $T_{cb}$ was 97 K, and $T_{cz}$ was 87 K). The authors of  
Ref.\ \onlinecite{clark} have assumed that $T_{cb}$ in this case  
reflects the behavior of intrinsic critical temperature of grains, which 
is not as sensitive to irradiation as that of the whole granular system.  
\par
In the years, following Ref.\ \onlinecite{clark}, the quality of HTSCs 
becomes much higher and the most of subsequent radiation studies were devoted 
to single-crystal (or, at least, nearly single crystal) HTSCs, which are 
usually far more homogeneous than polycrystalline samples, and for which, 
therefore, the phase coherence is not of decisive importance. It is certain,
however, that not only single crystal HTSC products may be (or will be) used 
in advanced technology. Therefore inhomogeneity effect in superconductivity 
of these compound are still of fundamental and tecnological importance. 
\par
It can be said with confidence that the observed decrease in $T_{c}$ and 
zero-resistance temperature $T_{cz}$ combined with simultaneous increase in 
width of resistive transition $\delta T_{c}$ in sample studied at low dose
($\Phi < 50$~MR) (Figs. 3 and 4) is quite expected for percolating granular 
system. Optimal current paths, which have ensured the measured $T_{c}$ value 
about 93 K before irradiation, have sure some ``weak'' links. These are 
the grain boundaries, which are strongly enough depleted with charge 
carriers and, therefore, are sensitive even to such small mean radiation 
damage as in this study (maximum $10^{-6}$ dpa). This leads to the observed 
decrease in the ``global'' $T_{c}$ and increase in $\delta T_{c}$ 
(Figs. 3 and 4).     
\par
The initial $T_c$ drop (which appears to be explicable) is followed 
by an increase in $T_c$ at higher dose $\Phi > 50$~MR (Fig. 3), and this is 
fairly surprising. This means that some concurrent mechanism, which causes 
an increase in $T_c$, comes into play. This mechanism shows itself only in 
the limited range of doses, since $T_c$ decreases again in the range above  
$\Phi > 120$~MR (Fig.~3). Somehow this mechanism should be also 
associated with an impact of $\gamma$-rays on intergrain Josephson coupling. 
There are known some cases of superconductivity enhancement in polycrystalline
HTSCs under irradiation. For example, an increase in $T_c$  after low-dose 
irradiation with Si ions was observed in polycrystalline YBCO\cite{mishra}.  
Explanations of the effect in Ref. \onlinecite{mishra} was not however 
connected with granular structure of sample. In Ref. \onlinecite{hasan}  
it was reported that $\gamma$-irradiation of granular 
Tl$_{2}$Ba$_{2}$Ca$_{2}$Cu$_{3}$O$_{10}$
at room temperature causes the increase of critical current. This is 
believed to be an indication of strengthening of the intergrain Josephson 
coupling under $\gamma$-ray influence. 
\par
The known features of interaction of $\gamma$-rays with 
solids\cite{bethe,dienes} make it possible to suggest some mechanisms 
of $T_c$ increasing in granular HTSCs at low enough $\gamma$-ray doses. 
In doing so two main points should be considered: (i) the nature of grain 
boundaries in HTSCs, (ii) the ionizing influence of $\gamma$-rays.
From the literature data (see Refs. \onlinecite{boyko,stolb,gilabert} and 
Refs. therein) it can be concluded that grain boundaries in HTSCs are 
disordered and oxygen deficient regions. A typical grain-boundary width 
in YBCO is about 2~nm\cite{boyko,stolb} (what encompasses about 10 atomic 
distances). It is known that $T_c$ of oxygen deficient YBCO compounds 
increases considerably with illumination of visible light. It was found
that light leads to a change in doping (increasing in charge carrier density).
This photoexcited state is persistent up to 250-270 K\cite{stock}. 
The photodoping effect is negligible for optimally doped YBCO. All existing 
models of the effect are based on the suggestion that sample illumination 
generates electron-hole pairs in the CuO$_2$ planes. Electrons are 
transferred to the CuO chains and trapped there, while holes remain 
mobile in the CuO$_2$ planes. 
\par
The most important effect of $\gamma$-rays in solids is ionization. 
In primary collision of $\gamma$-quantum with an atom of solid the Compton 
electron and photon with lesser energy are produced. The energy of 
secondary photon is dissipated in causing further ionization. The same is 
true for the most of the energy of Compton electrons (only occasionally they 
displace atoms by elastic collision). Therefore, the number of ionized atoms 
at $\gamma$-irradiation is by several orders of magnitude larger than 
the number of displaced atoms.  In this connection it must not be ruled out 
that effects like the above-mentioned photodoping by visible 
light\cite{gilabert,stock} take place in $\gamma$-irradiated bulk samples 
as well. Of course, there is a significant difference in the energy of 
$\gamma$-ray and visible light photons. The visible light can eject only 
electrons from outer orbitals. These low energy electrons are trapped in 
nearby sites like CuO chains. By contrast, $\gamma$-quantum can struck out 
any electron from an atom. But in this case the ejected elecrons have much 
higher energy and, therefore, can be brought far away from initial place.
If electrons are ejected from CuO$_2$ planes in oxygen depleted grain 
boundary regions, this should increase the charge carrier density in the 
planes and, therefore, enhance the Josephson coupling between adjacent 
grains. If this takes place in some of ``weak'' links in optimal current 
path, this will result in increasing of measured $T_c$. Of course, the 
recombination of the electrons with excess holes should bring the charge 
carrier density in grain boundaries to an initial level. But some of these 
electrons should be sure trapped in remote defect places of crystal lattice 
like impurities, lattice distortions regions, including grain boundaries 
as well. If room temperature is not high enough for the electron-hole 
recombination of trapped electrons, this can lead to the observed effect 
of $T_c$ enhancement in some intermediate dose range. At higher doses, 
however, influence of radiation defects (displaced atoms) in grain 
boundaries overpowers the ionizing effect of $\gamma$-rays that results in 
$T_c$ decreasing. Needless to say that our explanation should not be 
considered as a conclusive. It needs an independent theoretical and 
experimental confirmation.

\section{Conclusion}
We have found non-monotonic behavior of $T_c$ in optimally doped polycrystalline 
YBCO with increasing $\gamma$-ray irradiation dose. This result can be 
explained primarily by the influence of $\gamma$-rays on intergrain 
Josephson coupling. The two kinds of such influence are discussed. The 
first is the producing of displaced atoms, and the second is the ionizing
influence of $\gamma$-rays.

\begin{figure}
\caption{Temperature dependence of the resistivity of 
non-irradiated YBa$_{2}$Cu$_{3}$O$_{7-\delta}$ sample.}
\label{Fig.1}
\end{figure}

\begin{figure}
\caption{
The dependences $\rho(T)$ of sample studied in the temperature range of 
superconducting transition for different doses $\Phi$.  
}
\label{Fig.2}
\end{figure}

\begin{figure}
\caption{
The changes in $T_{c}$ and $T_{cz}$ with  $\gamma$-ray dose. The solid lines
are guides to the eye.
}
\label{Fig.3}
\end{figure}

\begin{figure}
\caption{
The change in $\delta T_{c} = T_{c} - T_{cz}$ with $\gamma$-ray dose. 
The solid line is a guide to the eye.
}
\label{Fig.4}
\end{figure}

\widetext
\begin{table}
\caption{Calculated values of the effective cross section for atomic 
displacement  by $\gamma$-rays from $^{60}$Co source through the Compton 
effect, $\sigma^{\gamma}_{c} = R_{d}^{\beta}/(\Phi_{\gamma}n_{0}$), 
at different values of threshold energy $E_{d}$ for ions in 
YBa$_{2}$Cu$_{3}$O$_{7-\delta}$ (in units $10^{-24}$~cm$^{2}$). Additional 
comments to this table are given in the text of this paper.}
\begin{tabular}{cccccc}
 &\multicolumn{4}{c}{$E_{d}$ (eV)}\\
 Ion&10&15&20&25&30\\ \tableline
 Y&1.09&0.335&0.116&0.041&0.014\\
 Ba&0.69&0.14&0.028&0.0045&$\approx 0.001$\\
 Cu&1.29&0.479&0.208&0.097&0.047\\
 O&1.13&0.58&0.349&0.23&0.16\\
 \end{tabular}
 \label{table1}
 \end{table}

\end{document}